\documentclass[showpacs,amsmath,amssymb,prd]{revtex4}
\usepackage[dvips]{graphicx} \newcommand{\tr}{\operatorname{Tr}}
\begin{document}

\title{Precise bounds on the Higgs boson mass}

\author{P.~Kielanowski} \email{kiel@physics.utexas.edu}
\affiliation{Departamento de F\'{\i}sica, Centro de Investigaci\'{o}n
  y Estudios Avanzados del IPN, Mexico}

\author{S.R.~Ju\'{a}rez~W.}  \email{rebeca@esfm.ipn.mx}
\altaffiliation{on sabbatical leave at Departamento de F\'{\i}sica,
  Centro de Investigaci\'{o}n y Estudios Avanzados del IPN, Mexico}

\affiliation{Departamento de F\'{\i}sica, Escuela Superior de
  F\'{\i}sica y Matem\'{a}ticas, IPN, Mexico}

\begin{abstract}
  We study the renormalization group evolution of the Higgs quartic
  coupling $\lambda_{H}$ and the Higgs mass $m_{H}$ in the Standard
  Model. The one loop equation for $\lambda_{H}$ is non linear and it
  is of the Riccati type which we numerically and analytically solve
  in the energy range $[m_{t},E_{GU}]$ where $m_{t}$ is the mass of
  the top quark and $E_{GU}=10^{14}$~GeV. We find that depending on
  the value of $\lambda_{H}(m_{t})$ the solution for $\lambda_{H}(E)$
  may have singularities or zeros and become negative in the former
  energy range so the ultra violet cut off of the standard model
  should be below the energy where the zero or singularity of
  $\lambda_{H}$ occurs.  We find that for
  $0.369\leq\lambda_{H}(m_{t})\leq0.613$ the Standard Model is valid
  in the whole range $[m_{t},E_{GU}]$. We consider two cases of the
  Higgs mass relation to the parameters of the standard model: (a)~the
  effective potential method and~(b) the tree level mass relations.
  The limits for $\lambda_{H}(m_{t})$ correspond to the following
  Higgs mass relation $150\leq m_{H}\lessapprox 193$~GeV. We also plot
  the dependence of the ultra violet cut off on the value of the Higgs
  mass. We analyze the evolution of the vacuum expectation value of
  the Higgs field and show that it depends on the value of the Higgs
  mass.  The pattern of the energy behavior of the VEV is different
  for the cases (a)~and~(b).  The behavior of $\lambda_{H}(E)$,
  $m_{H}(E)$ and $v(E)$ indicates the existence of a phase transition
  in the standard model. For the effective potential this phase
  transition occurs at the mass range $m_{H}\approx 180$~GeV and for
  the tree level mass relations at $m_{H}\approx 168$~GeV.
\end{abstract}

\maketitle

\section{Introduction}

The Standard Model (SM) provides a very precise description of all the
present elementary particle data~\cite{ref1}. On the other hand it has
relatively many free parameters ($\sim 19$) what is rather
unsatisfactory from the fundamental point of view. The idea of Grand
Unification (GU)~\cite{ref2} is to look for additional symmetries in
the SM at very high energies. The most notable sign of the presence of
GU is the (approximate) convergence of the three gauge couplings to
one common value at the energies $ 10^{14}-10^{15}$~GeV. This allows
to substitute the gauge group $SU(3)\times SU(2)\times U(1)$ of the SM
by a larger group and to reduce the number of gauge couplings to only
one.

The main tool of the GU models are the Renormalization Group Equations
(RGE)~\cite{ref3} that relate various observables (like couplings or
masses) at different energies and also allow the study of their
asymptotic behavior.

In the perturbative quantum field theory the RGE are differential
equations for the observables which are obtained from the condition
that the $S$-matrix elements do not depend on the renormalization
scheme or renormalization point. The right hand side of the RGE is an
infinite series expanded according to the number of loops. Most of the
numerical and analytical results for the RGE \cite{ref4} are to the
order of one loop only possibly with a partial inclusion of two loops,
while the RGE for most of the observables have been given for the SM
and its extensions up to two loops~\cite{ref4,ref5,ref5a}.

The right hand side of the RGE is constructed from the following terms
\begin{equation}
  g_{i}^{2},\quad
  y_{u}^{\phantom{\dagger}}y_{u}^{\dagger },\quad
  y_{d}^{\phantom{\dagger}}y_{d}^{\dagger},\quad
  y_{l}^{\phantom{\dagger}}y_{l}^{\dagger },\quad
  y_{\nu }^{\phantom{\dagger}}y_{\nu }^{\dagger},\quad
  \lambda _{H},
\end{equation}
where the $g_{i}^{\phantom{\dagger}}$'s are the gauge couplings,
$y_{u}^{\phantom{\dagger}}$, $y_{d}^{\phantom{\dagger}}$,
$y_{l}^{\phantom{\dagger}}$, $y_{\nu}^{\phantom{\dagger}}$, are the
Yukawa couplings of the up and down quarks, charged leptons and
neutrinos, respectively and $\lambda_{H}$ is the Higgs quartic
coupling constant. The RGE form a set of non-linear coupled
differential equations and even at the one loop order there exist only
approximate or numerical solutions \cite{ref4,ref6}.

The one loop RGE for the best measured observables $g_{i}$'s, quark
and lepton Yukawa couplings and the Cabibbo-Kobayashi-Maskawa (CKM)
matrix are independent of the Higgs quartic coupling. This allows to
derive the running of those observables at the lowest order without
the knowledge of the $\lambda_{H}.$ On the other hand, at the two loop
level, the quartic coupling $\lambda_{H}$ appears in the RGE for many
observables like quark masses or the CKM matrix and has important
influence on their behavior and cannot be neglected.

The one loop equation for $\lambda_{H}$ is also non linear and has
been used to obtain the limits on the Higgs mass from the triviality
of the $\lambda\phi^{4}$ theory and the existence of the Landau pole.
This equation has been also considered in Refs.~\cite{ref7,ref8,ref8a}
to study the dependence of the Higgs mass and the UV cut off on the
energy and it was solved for the simplified case when the gauge
couplings and the top quark Yukawa coupling are constant.

In this paper we study the one loop equation for $\lambda_{H}$ without
any simplifying assumptions in the energy range, starting at the top
quark mass $m_{t}$. We find that the equation is of the Riccati type
and we solve this equation explicitly. We find that for the values of
$\lambda_{H}$ at the top quark mass, $\lambda_{H}(m_{t})\geq0.528$,
the function $\lambda_{H}(E)$ has a Landau singularity. For the
values of $\lambda_{H}(m_{t})\leq 0.386$ there is no Landau pole below
the energies $E_{GU}$ and the solution $\lambda_{H}(E)$ passes through
zero and then becomes negative. This means that for the latter values
of $\lambda_{H}(m_{t})$ the theory becomes unstable and the UV cutoff
should appear below the energy value corresponding to the zero of
$\lambda_{H}(E)$. As is well known the coupling $\lambda_{H}$ is
related to the Higgs mass so our results are presented in terms of the
Higgs mass.

We also study the one loop RGE for the Higgs mass $m_{H}$. We consider
two cases here. The first one is where the Higgs mass is obtained from
the Higgs field effective potential and in the other one the tree
level relation of the Higgs mass with the parameters of the SM
Lagrangian is used.  We explicitly solve theses equations for both
cases and use them to find the UV cut off of the standard model and
show the evolution of the Higgs mass in the range $120\leq
m_{H}\leq500$~GeV.  We next use the evolution of the $m_{H}$ and
$\lambda_{H}$ to find the evolution of the vacuum expectation value of
the Higgs field. All these results show that there is a sharp
transition in the behavior of the standard model at the Higgs mass
$m_{H}\approx 180$~GeV for the case of the effective potential and at
$m_{H}\approx 168$~GeV for the tree level relations.  Additionally we
show that for the case of the effective potential for the Higgs mass
in the range $150\leq m_{H}\leq 193$~GeV the standard model is valid
up to the GU energy $\sim10^{14}$~GeV. For the case of the tree level
mass relations the former limit is $150\leq m_{H}\leq 194$~GeV.

\section{One and Two Loops Renormalization Group Equations}

The two loop RGE are the following
\begin{subequations}
\label{eq2}
\begin{equation}
  \frac{dg_{l}}{dt}
  =\frac{1}{(4\pi)^{2}}b_{l} g_{l}^{3}
  -\frac{1}{(4\pi)^{4}}G_{l}g_{l}^{3},
  \label{0.1}
\end{equation}
\begin{equation}
  \frac{dy_{u,d,e,\nu }}{dt}=
  \left[\frac{1}{(4\pi )^{2}}\beta _{u,d,e,\nu}^{(1)}
    +\frac{1}{(4\pi )^{4}}\beta _{u,d,e,\nu }^{(2)}\right] y_{u,d,e,\nu },
\label{0.2}
\end{equation}
\begin{equation}
  \frac{d\lambda_{H}}{dt}
  =\left[\frac{1}{(4\pi )^{2}}\beta_{\lambda}^{(1)}
    +\frac{1}{(4\pi)^{4}}\beta_{\lambda }^{(2)}\right],
\label{0.4}
\end{equation}
\begin{equation}
  \frac{d m^{2}}{dt}=
  \left[\frac{1}{(4\pi)^{2}}\beta^{(1)}_{m^{2}}
    +\frac{1}{(4\pi)^{4}}\beta^{(2)}_{m^{2}}\right].  \label{0.3}
\end{equation}
\end{subequations}
where $t=\ln(E/m_{t})$, $E$ is the energy and $m_{t}=174.1$~GeV is the
top quark mass and the Higgs potential is
$m^{2}\phi^{\dagger}\phi+(\lambda_{H}/2)(\phi^{\dagger}\phi)^{2}$.
The constants $b_{l}$ depend on the model and $G_{l}$,
$\beta_{u,d,e,\nu }^{(1)}$, $\beta_{u,d,e,\nu }^{(2)}$,
$\beta_{\lambda}^{(1)}$, $\beta_{\lambda}^{(2)}$, $\beta^{(1)}_{m^{2}}$,
$\beta^{(2)}_{m^{2}}$ are functions of the standard model couplings and the
squares of the Yukawa couplings $H_{u,d,e,\nu}^{(1)}=y_{u,d,e,\nu
}y_{u,d,e,\nu}^{\dagger}$, (for the definition of these functions and
constants see~\cite{ref4} or~\cite{ref6}).

In the previous papers~\cite{ref6} we have discussed a consistent
approximation scheme for the solution of the RGE that was based on the
expansion of the solutions in terms of the powers of $\lambda$, where
$\lambda\simeq 0.22$ is the absolute value of the
$\left|V_{us}\right|$ element of the CKM matrix.

In such an approximation the lowest order RGE have the following
form~\cite{ref5,ref5a}~\footnote{We neglect the leptonic part of the RGE
  because it decouples from the quark sector in the considered
  approximation.}
\begin{subequations}
\label{eq1}
\begin{equation}
  \frac{dg_{i}}{dt}=\frac{1}{(4\pi )^{2}}b_{i}g_{i}^{3},\quad i=1,2,3,
  \label{1}
\end{equation}
\begin{equation}
  \frac{dy_{u}}{dt}=
  \frac{1}{(4\pi)^{2}}\left\{\alpha_{1}^{u}(t)
  +\alpha_{2}^{u}y_{u}^{\phantom{\dagger}}y_{u}^{\dagger}
  +\alpha _{3}^{u}
  \tr(y_{u}^{\phantom{\dagger}}y_{u}^{\dagger})\right\}y_{u},
  \label{2}
\end{equation}
\begin{equation}
  \frac{dy_{d}}{dt}
  =\frac{1}{(4\pi)^{2}}\left\{\alpha _{1}^{d}(t)
    +\alpha_{2}^{d}y_{u}^{\phantom{\dagger}}y_{u}^{\dagger }
    +\alpha _{3}^{d}
    \tr(y_{u}^{\phantom{\dagger}}y_{u}^{\dagger})\right\}y_{d},
  \label{3}
\end{equation}
\begin{equation}
  \frac{d\lambda _{H}}{dt}
  =\frac{12}{(4\pi )^{2}}\left\{\lambda_{H}^{2}
    +\left[\tr(y_{u}^{\phantom{\dagger}}y_{u}^{\dagger })
      -\frac{3}{4}\left(\frac{1}{5}g_{1}^{2}+g_{2}^{2}\right)\right]
    \lambda_{H}
    +\frac{3}{16}
    \left(\frac{3}{25}g_{1}^{4}+\frac{2}{5}g_{1}^{2}g_{2}^{2}+g_{2}^{4}\right)
    -\tr(y_{u}^{\phantom{\dagger}}y_{u}^{\dagger })^{2}\right\},
  \label{5}
\end{equation}
\begin{equation}
  \frac{d\ln m^{2}}{dt}
  =\frac{1}{(4\pi)^{2}}
  \left\{6\lambda_{H} 
    +6\tr(y_{u}^{\phantom{\dagger}}y_{u}^{\dagger })
  -\frac{9}{2}\left(\frac{1}{5}g_{1}^{2}+g_{2}^{2}\right)\right\}.
  \label{4}
\end{equation}
\end{subequations}
The constants $b_{i}$ and $\alpha$'s in Eqs.~(\ref{eq1}) are equal
\begin{alignat*}{4}
  &(b_{1},b_{2},b_{3})=(\frac{41}{10},-\frac{19}{6},-7)\\
  \alpha_{1}^{u}(t) =&-(\frac{17}{20}g_{1}^{2}+\frac{9}{4}
  g_{2}^{2}+8g_{3}^{2}),&\quad& \alpha_{2}^{u}&=\frac{3}{2}b,&\quad&
  \alpha_{3}^{u}&=3, \\
  \alpha _{1}^{d}(t) =&-(\frac{1}{4}g_{1}^{2}+\frac{9}{4}
  g_{2}^{2}+8g_{3}^{2}),&& \alpha_{2}^{d}&=\frac{3}{2}c,&&
  \alpha_{3}^{d}&=3a, \\
  &(a,b,c)=(1,1,-1).
\end{alignat*}
Eqs.~(\ref{1})-(\ref{3}) can be \emph{explicitly} solved and the most
important results and properties of these solutions are~\cite{ref6}
\begin{enumerate}
\item $g_{i}$'s, $y_{u}$ and $y_{d}$ are all regular functions of
  energy in the range $[m_{t},E_{GU}]$.
\item The running of the gauge couplings $g_{l}(t)$ is
  ($t_{0}=\ln(E/m_{t})|_{E=m_{t}}=0$)
  \begin{equation}
    \label{6}
    \left(g_{i}(t)\right)^{2}
    =\frac{\left( g_{i}(t_{0})\right) ^{2}}{1
      -\frac{2}{(4\pi )^{2}}\left( g_{i}(t_{0})\right)^{2}b_{i}(t-t_{0})}.
  \end{equation}
\item The running of the up quark Higgs couplings $y_{u}(t)$ has the
  following property
  \begin{equation}
    \label{6a}
    \tr(y_{u}^{\phantom{\dagger}}y_{u}^{\dagger })=Y^{2}_{t}(t)=
    \frac{Y^{2}_{t}(t_{0})r(t)}{1
    -\frac{2(\alpha_{2}^{u}+\alpha _{3}^{u})}
    {(4\pi)^{2}}Y^{2}_{t}(t_{0})\int_{t_{0}}^{t}r(\tau)d\tau}
  \end{equation}
  where $Y_{t}$ is the largest eigenvalue of the up quark Higgs
  coupling matrix $y_{u}$ and $r(t)
  =\exp\left((2/(4\pi)^{2})\int_{t_{0}}^{t}\alpha_{1}^{u}(\tau) d\tau
  \right)=\prod_{k=1}^{3}[g^{2}_{k}(t_{0})/g^{2}_{k}(t)]^{c_{k}/b_{k}}$,
  $c_{k}=(17/20,9/4,8)$.
\end{enumerate}
Using Eqs.~\eqref{6} and~\eqref{6a} as input into Eq.~\eqref{5} we
obtain the uncoupled differential equation for the quartic coupling
constant $\lambda_{H}$.

Eq.~\eqref{5} for $\lambda_{H}$ has been considered earlier by various
authors~\cite{ref7,ref8,ref8a} but in all these papers the effects of
the running of the gauge couplings and of $Y_{t}^{2}$ have not been
considered.  The importance of $\lambda_{H}$ for the evolution of
other observables comes from the fact that $\lambda_{H}$ appears at
the two loop order in the RGE for $y_{u}$ and $y_{d}$ and at the one
loop order for $m_{H}$.

\section{One loop equation for $\lambda _{H}$}

The one loop equation for $\lambda_{H}$ given in Eq.~(\ref{5}) is
rewritten in the form
\begin{equation}
  \label{eq9a}
  \frac{d\lambda_{H}}{dt}=f_{0}(t)+f_{1}(t)\lambda_{H}
  +f_{2}(t)\lambda_{H}^{2},
\end{equation}
where the definition of the functions $f_{i}(t)$ can be deduced from
Eq.~\eqref{5}.  This equation is of the Riccati type~\cite{ref9}. The
behavior of the gauge coupling $g_{i}$'s is given in Eq.~(\ref{6}) and
the explicit energy dependence of
$\tr(y_{u}^{\phantom{\dagger}}y_{u}^{\dagger})$ is given in
Eq.~\eqref{6a}. As discussed before the $g_{i}$'s and
$\tr(y_{u}^{\phantom {\dagger}}y_{u}^{\dagger})$, as functions of
energy, have no singularities in the range $[m_{t},E_{GU}]$.  On the
other hand the solutions of the Riccati's equations can become
singular even if the coefficients of the equation are smooth and
regular functions.

The solution of Eq.~\eqref{eq9a} is obtained by substituting the
$\lambda_{H}$ by the following expression containing the auxiliary
function $W(t)$
\begin{equation}
  \label{eq9b}
  \lambda_{H}(t)=-\frac{1}{f_{2}(t)}\frac{W^{\prime}(t)}{W(t)}
\end{equation}
which fulfills the linear second order differential equation
\begin{equation}
  W^{\prime \prime }
  -\left(\frac{f_{2}^{\prime}(t)}{f_{2}(t)}+f_{1}(t)\right)
  W^{\prime} +f_{0}(t)f_{2}(t)W =0.  \label{15}
\end{equation}
Any solution of Eq.~(\ref{15}) generates the solutions of
Eq.~\eqref{eq9a}. Eq.~(\ref{15}) is of the Frobenius type \cite{ref10}
and the solution $W(t)$ is a \textit{regular} function of the energy
$t$ in the region where the coefficients of Eq.~(\ref{15}) are
\textit{regular}. One can look for the solutions of this equation in
terms of an infinite series. We look for the two solutions of this
equation with the following properties
\begin{align}
  \left.W_{1}(t)\right|_{t_{0}} &=1,\quad\left. W_{1}^{\prime
    }(t)\right|_{t_{0}}=0,
  \nonumber \\
  \left.W_{2}(t)\right|_{t_{0}} &=0,\quad\left. W_{2}^{\prime
    }(t)\right|_{t_{0}}=1.
  \label{16}
\end{align}
The solution of~\eqref{eq9a} for ${\lambda }_{H}$ in terms of the
functions $W_{1}(t) $ and $W_{2}(t)$ has the following form (note that
$f_{2}(t)=12/(4\pi)^{2}$)
\begin{equation}
  \lambda_{H}(t) =-\frac{(4\pi )^{2}}{12}
  \frac{W_{1}^{\prime}(t)-\frac{12}{(4\pi )^{2}}\lambda_{H}(t_{0})
    W_{2}^{\prime }(t)}
  {W_{1}(t) -\frac{12}{(4\pi )^{2}} \lambda_{H}(t_{0}) W_{2}(t)}.
  \label{17}
\end{equation}
The most important property of the solution (\ref{17}) is that the
singularities of the solution $\lambda_{H}(t)$ are determined from the
zeros of the denominator
\begin{equation}
  W_{1}(t) -\frac{12}{(4\pi )^{2}}\lambda_{H}(t_{0})
  W_{2}(t) =0  \label{19}
\end{equation}
and the zeros of $\lambda_{H}(t)$ are determined from the zeros of the
numerator
\begin{equation}
  W_{1}^{\prime}(t) -\frac{12}{(4\pi )^{2}}\lambda_{H}(t_{0})
  W_{2}^{\prime}(t) =0  \label{19a}
\end{equation}
then one can precisely determine the position of the singularities and
zeros and their dependence on the initial value of the Higgs quartic
coupling $\lambda_{H}(t_{0})$. The detailed discussion of the
solutions is given in the next section.

\section{Running of $\lambda_{H}$}

\begin{figure}[b]
  \centering
  \includegraphics[width=0.8\textwidth]{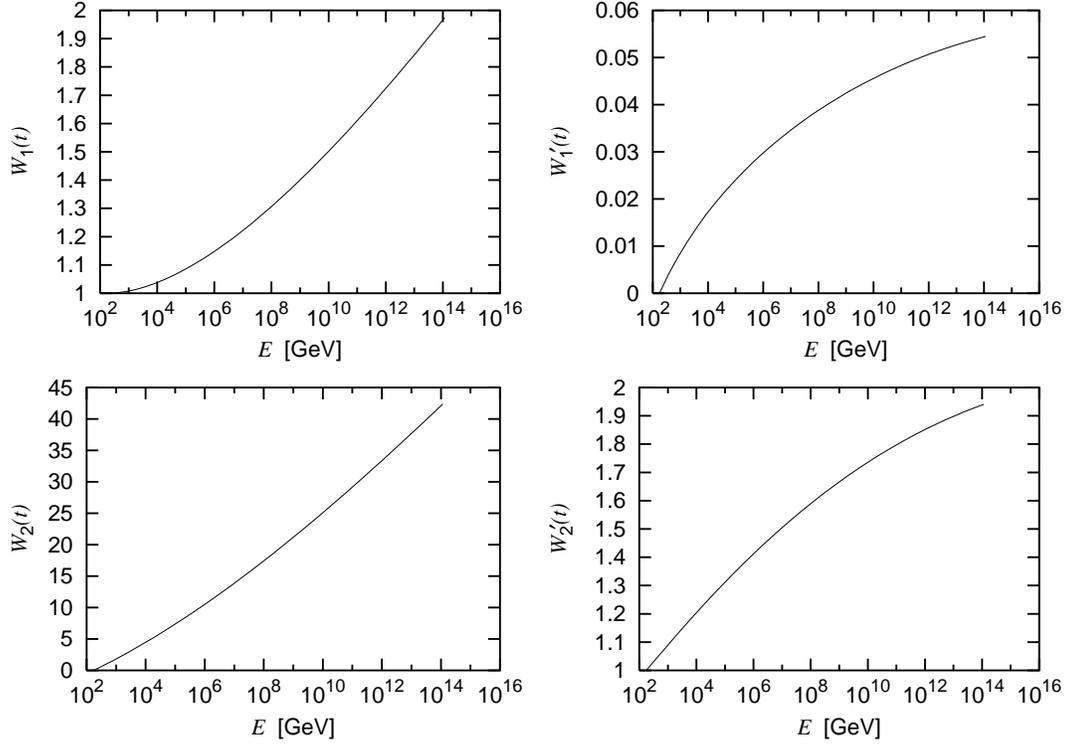}
\caption{\label{fig1}
  The solutions of~Eq~(\ref{15}) and their derivatives.}
\end{figure}

In this section we will discuss the explicit solutions of
Eqs.~(\ref{15}) and (\ref{5}). Let us start with Eq.~(\ref{15}).  The
form of the functions $-({f_{2}^{\prime}(t)}/{f_{2}(t)}+f_{1}(t))$ and
$f_{0}(t)f_{2}(t)$ is too complicated to be able to solve
Eq.~(\ref{15}) explicitly. To find the solution of this equation we
use the fact that they are smooth functions of energy so we
approximate these two functions in the energy range
$\left[m_{t},E_{GU}\right]$ by the ratio of two polynomials. These
functions perfectly approximate both coefficients in Eq.~\eqref{15} in
the whole energy range and this allows to find the solution of
Eq.~(\ref{15}) in terms of a power series of the variable
$t$~\footnote{We applied two methods of the solution of
  Eq.~\eqref{15}, one based on the explicit power expansions and the
  other using Mathematica capabilities of the numerical solution of
  the differential equations. Both approaches agree perfectly.}. In
Fig.~\ref{fig1} we show the dependence on the energy of the two
solutions of Eq.~\eqref{15} and their derivatives~\footnote{In the
  numerical calculations we use the following initial values of the
  parameters: $g_{1}(t_{0})=0.4459$, $g_{2}(t_{0})=0.62947$,
  $g_{3}(t_{0})=1.2136$, $m_{t}(t_{0})=174.1$ and $v(t_{0})=174.1$}.
As expected they are smooth functions of~$t$.

From Eq.~\eqref{17} we find now the dependence of $\lambda_{H}(t)$ on
the energy $t$ and important properties of its behavior.  It is the
most interesting to investigate how $\lambda_{H}(t)$ depends on the
initial values of $\lambda_{H}(t_{0})$ and to find out the range of
validity of the SM. As discussed earlier, for the SM to be valid
$\lambda_{H}(t)$ must be positive and cannot be singular. Since
$\lambda_{H}(t_{0})>0$, it means that the SM is valid for energies
between $m_{t}$ and the zero or singularity of $\lambda_{H}(t)$ which
can be determined from Eqs.~\eqref{19} and \eqref{19a}.

\begin{figure}[!tb]
  \centering \includegraphics[height=0.4\textheight]{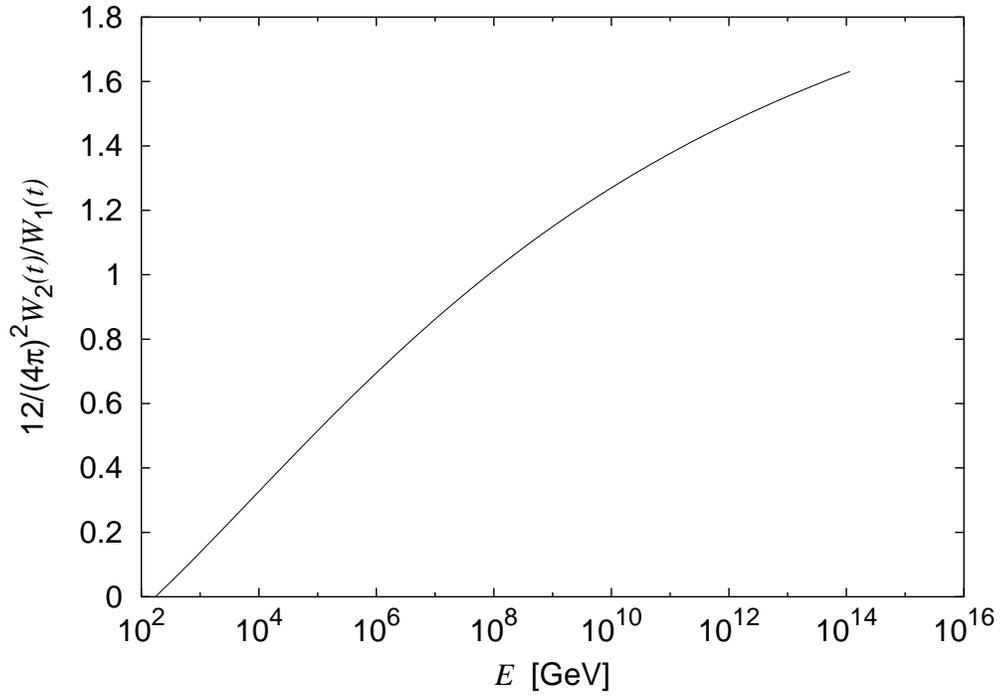}
\caption{\label{fig2}
  The ratio of the solutions $(12/(4\pi)^{2})W_{2}(t)/W_{1}(t)$
  of~Eq(~(\ref{15})). This ratio determines the value of $t$ at which
  $1/\lambda_{H}(t)$ vanishes, i.e.\ $\lambda_{H}(t)$ and $m_{H}^{4}(t)$
  have a pole and $v^{4}(t)$ has a zero.}
\end{figure}

\begin{figure}[!bt]
  \centering \includegraphics[height=0.4\textheight]{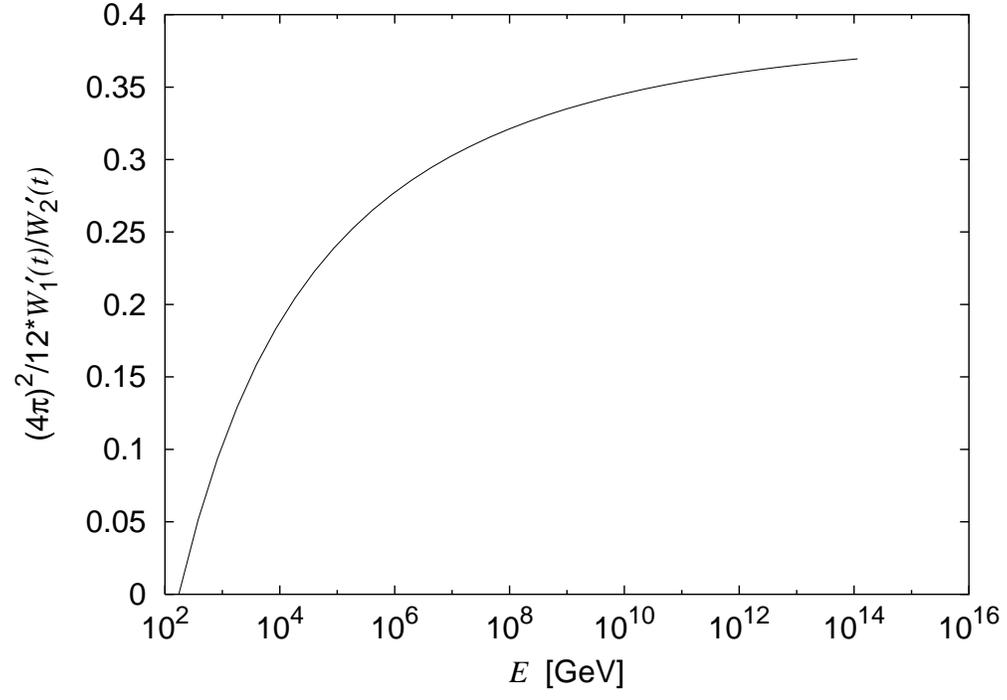}
\caption{\label{fig3}
  The ratio of the derivatives
  $((4\pi)^{2}/12)W_{1}^{\prime}(t)/W_{2}^{\prime}(t)$ of the
  solutions of~Eq(~(\ref{15})). This ratio determines the position $t$
  at which $\lambda_{H}(t)$ has a zero and $v^{2}(t)$ has a pole.}
\end{figure}

Let us first consider the singularity (a simple pole) of
$\lambda_{H}(t)$.  For this purpose we plot in Fig.~\ref{fig2} the
ratio of the two solutions $(12/(4\pi)^{2})W_{2}(t)/W_{1}(t)$ from
which we can determine the value of $t$ for which the pole occurs
depending on the value of $\lambda_{H}(t_{0})$. If we impose the
condition that $\lambda_{H}(t)$ is regular in the whole range of the
energies $[m_{t},E_{GU}]$ then the value of $1/\lambda_{H}(t_{0})$
should lie above the curve in Fig.~\ref{fig2} what gives the following
condition
\begin{equation}
  \lambda_{H}(t_{0})\leq 0.613.
  \label{20}
\end{equation}
For the SM to be valid the quartic coupling $\lambda_{H}(t)$ should
not become negative. We use Eq.~\eqref{19a} to find the first zero of
$\lambda_{H}(t)$. In Fig.~\ref{fig3} we have plotted
$((4\pi)^{2}/12)W_{1}^{\prime}(t)/W_{2}^{\prime}(t)$ which determines
at which energy in $t$ occurs the first zero of $\lambda_{H}(t)$
depending on the value of $\lambda_{H}(t_{0})$.  Now from the
condition that $\lambda_{H}(t)$ should not have zeros in the whole
range of the energies $[m_{t},E_{GU}]$ we obtain
\begin{equation}
    \lambda_{H}(t_{0})\geq0.369.
  \label{21}
\end{equation}
We thus see that the consistency of the SM in the range of the
energies up to the grand unification energy $E_{GU}$ permits a very
narrow band on the admissible values of the $\lambda_{H}(t_{0}).$
\begin{equation}
  \label{eq21a}
  0.369 \leq \lambda_{H}(t_{0}) \leq 0.613.
\end{equation}
In the next section we discuss the implications of the running of
$\lambda_{H}$ for the Higgs mass and the VEV of the Higgs field.

\section{Running of the Higgs mass and VEV}

The most interesting predictions from the one loop \emph{exact}
solution for $\lambda_{H}$ can be obtained for the Higgs boson mass.
This problem has received earlier a wide attention~\cite{sher} and we
will discuss here the method based on the Higgs effective
potential~\cite{casas1,casas2}. Recently there also appeared a series
of papers~\cite{kalmykov} in which the full two loop \emph{analytical}
analysis of the pole masses of the gauge bosons has been made. It has
been shown there that the tree level relations between the pole masses
and the couplings of the standard model are valid up to two loops.

We will compare the predictions of these two approaches for the
running of the Higgs field mass and VEV.

\subsection{Effective potential method}

The square of the Higgs boson mass is defined as the second derivative
of the effective potential taken at the minimum of the
potential~\cite{carena}. We will use the following form of the
effective potential~\cite[Eq.~(13)]{ref5}
\begin{equation}
  \label{veff}
  V_{\text{eff}}=m^{2}(t)Z^{2}(t)\phi^{\dagger}\phi+
  \frac{1}{2}\lambda_{H}(t)Z^{4}(t)(\phi^{\dagger}\phi)^2
\end{equation}
where $Z(t)$ is the renormalization factor of the Higgs field that
fulfills the RG equation~\cite[Eq.~(10)]{casas1}
\begin{equation}
  \label{zequation}
  \frac{d\ln Z}{dt}=-\gamma_{\phi}(g_{i}^{2},Y_{t}^{2})=
  \frac{3}{(4\pi)^{2}}\left(\frac{3}{20}g_{1}^{2}+\frac{3}{4}g_{2}^{2}
  -Y_{t}^{2}\right)
\end{equation}
which has the solution
\begin{equation}
  \label{zsolution}
  Z(t)=h_{m}^{-3}(t)\left(\frac{g_{1}(t)}{g_{1}(t_{0})}
  \right)^{\frac{9}{20b_{1}}}
  \left(\frac{g_{2}(t)}{g_{2}(t_{0})}
  \right)^{\frac{9}{4b_{2}}}
\end{equation}
and $h_{m}(t)$ is equal~\cite{ref6}
\begin{equation}
  \label{hm}
  h_{m}=\exp\left(\frac{1}{(4\pi)^{2}}
    \int_{t_{0}}^{t}
    \tr( y_{u}^{\phantom{\dagger}}y_{u}^{\dagger })dt\right).
\end{equation}

From Eq.~\eqref{veff} we obtain the following result for the physical
Higgs mass
\begin{equation}
  \label{higgsmass}
  m_{H}^{2}(t)=-2m^{2}(t)Z^{2}(t)
\end{equation}
and now combining Eqs.~\eqref{4} and~\eqref{zequation} we obtain the
following RG equation for the physical Higgs mass
\begin{equation}
  \label{hgmasseq}
  \frac{d\ln m_{H}^{2}}{dt}=\frac{6}{(4\pi)^{2}}\lambda_{H}.
\end{equation}
which has the following simple solution
\begin{equation}
  \label{eq21d1}
  m_{H}^{2}(t)= \frac{m_{H}^{2}(t_{0})}
  {\left(W_{1}(t)
      -\frac{12}{(4\pi)^{2}}\lambda_{H}(t_{0})W_{2}(t)\right)^{1/2}}
\end{equation}

The Higgs field vacuum expectation value can also be calculated from
the effective potential~\eqref{veff} and is equal
\begin{equation}
  \label{eq21d3}
  v^2(t)=\frac{m_{H}^{2}(t)}{2\lambda_{H}(t) Z^{4}(t)}
   =-m_{H}^{2}(t_{0})\frac{6}{(4\pi)^{2}}
  \frac{\left(W_{1}(t)
      -\frac{12}{(4\pi)^{2}}\lambda_{H}(t_{0})W_{2}(t)\right)^{1/2}}
  {W_{1}^{\prime}(t)
      -\frac{12}{(4\pi)^{2}}\lambda_{H}(t_{0})W_{2}^{\prime}(t)}
  \,\,h_{m}^{12}(t)
  \left(\frac{g_{1}(t_{0})}{g_{1}(t)}\right)^{\frac{9}{5b_{1}}}
  \left(\frac{g_{2}(t_{0})}{g_{2}(t)}\right)^{\frac{9}{b_{2}}}.
\end{equation}
We thus see that $m_{H}^{4}(t)$ and $\lambda_{H}(t)$ have a pole in
the same position.  $v^{2}(t)$ has a pole where $\lambda_{H}(t)$ has a
zero and $v^{2}(t)$ has a zero at the position of the Landau pole.

\subsection{Tree level relations}
In the case of the tree level relations the pole mass and the VEV of
the Higgs field are given by the following simple
relations~\cite{kalmykov}
\begin{equation}
  \label{kal1}
  m_{H}^{2}(t)=-2m^{2}(t),\quad
  v^{2}(t)=\frac{m_{H}^{2}(t)}{2\lambda_{H}(t)}. 
\end{equation}
From previous equation it follows that the RG equation for the Higgs
mass is the same as for the parameter $m^{2}$ of the Higgs potential
\begin{equation}
  \label{kal2}
   \frac{d\ln m^{2}_{H}}{dt}
  =\frac{1}{(4\pi)^{2}}
  \left\{6\lambda_{H} 
    +6\tr(y_{u}^{\phantom{\dagger}}y_{u}^{\dagger })
  -\frac{9}{2}\left(\frac{1}{5}g_{1}^{2}+g_{2}^{2}\right)\right\}.
\end{equation}
which has the following solution
\begin{equation}
  \label{kal3}
  m^{2}_{H}(t)= \frac{m_{H}^{2}(t_{0})}
  {\left(W_{1}(t)-\frac{12}{(4\pi)^{2}}
  \lambda_{H}(t_{0})W_{2}(t)\right)^{1/2}}
  h_{m}^{6}(t)\left(\frac{g_{1}(t_{0})}{g_{1}(t)}
  \right)^{\frac{9}{10b_{1}}}
  \left(\frac{g_{2}(t_{0})}{g_{2}(t)}
  \right)^{\frac{9}{2b_{2}}}
\end{equation}
and the VEV of the Higgs field is equal
\begin{equation}
  \label{kal4}
  v^{2}(t)=\frac{m_{H}^{2}(t)}{2\lambda_{H}(t)}
  =-m_{H}^{2}(t_{0})\frac{6}{(4\pi)^{2}}
  \frac{\left(W_{1}(t)
      -\frac{12}{(4\pi)^{2}}\lambda_{H}(t_{0})W_{2}(t)\right)^{1/2}}
  {W_{1}^{\prime}(t)
      -\frac{12}{(4\pi)^{2}}\lambda_{H}(t_{0})W_{2}^{\prime}(t)}
  \,\,h_{m}^{6}(t)
  \left(\frac{g_{1}(t_{0})}{g_{1}(t)}\right)^{\frac{9}{10b_{1}}}
  \left(\frac{g_{2}(t_{0})}{g_{2}(t)}\right)^{\frac{9}{2b_{2}}}.
\end{equation}
The analytical properties of the Higgs mass and the VEV of the Higgs
field are similar to the previous case.
\section{Conclusions}
\begin{figure}[!tb]
  \centering \includegraphics[height=0.8\textheight]{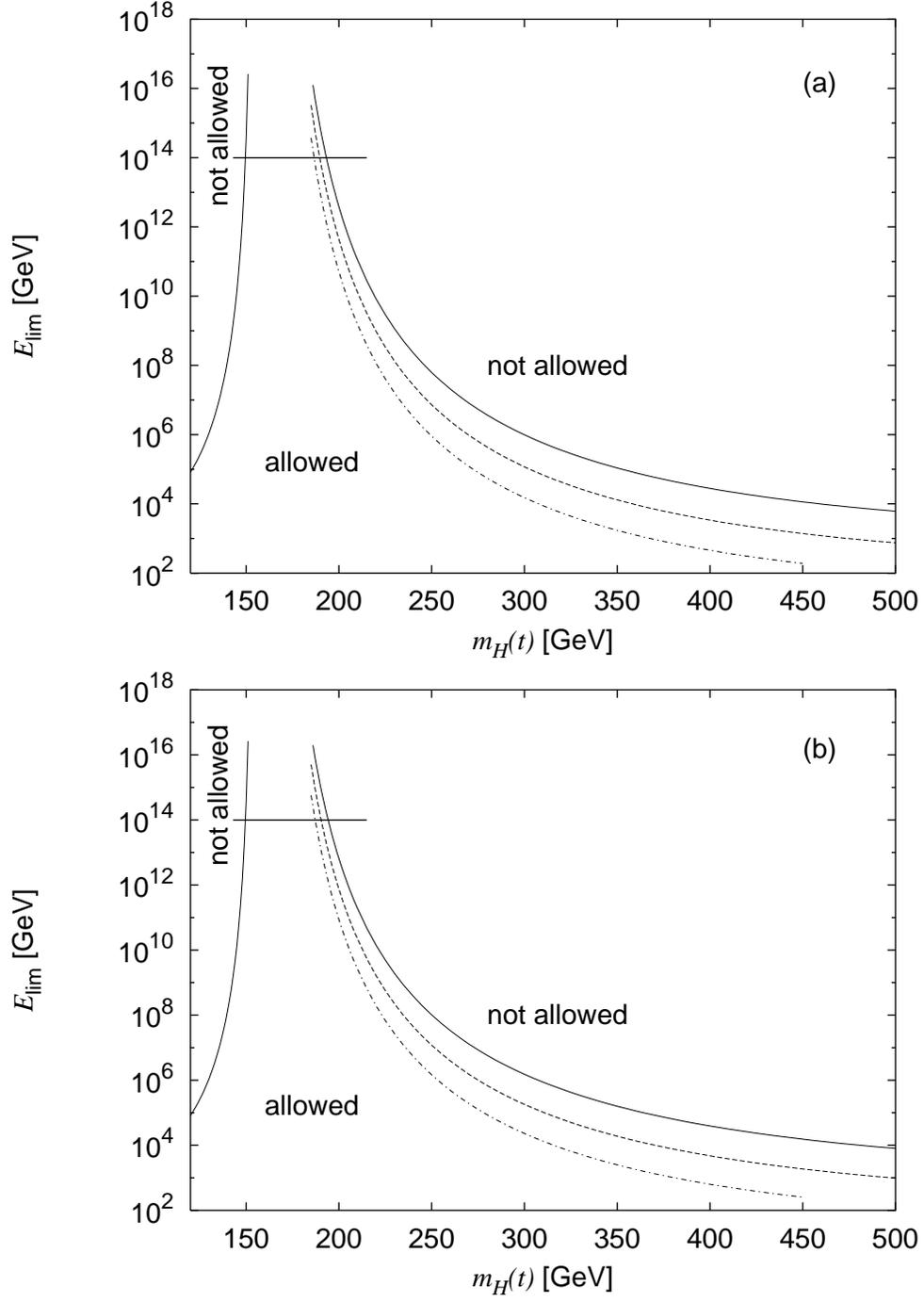}
\caption{\label{fig4}
  The plot of the energy $E_{\text{lim}}$ at which the SM breaks down
  as the function of the Higgs mass. Figure~(a) corresponds to the
  case of the effective potential and figure~(b) to the tree-level
  relation for the Higgs mass.  The curve on the left hand side is
  derived from the vanishing of $\lambda_{H}(t)$. The three curves on
  the right hand side are obtained from the condition that the
  perturbation expansion of the renormalization group equation for
  $\lambda_{H}$ breaks down: the continuous line follows from the
  position of the Landau pole, the middle curve corresponds to the
  value $\lambda_{H}=\lambda_{FP}/2$ and the lowest one to
  $\lambda_{H}=\lambda_{FP}/4$ ($\lambda_{FP}\approx12.1$ is the fixed
  point value of the two loop $\beta$ function for
  $\lambda_{H}$~\cite{ref8}). Both cases~(a) and~(b) give very similar
  form for the Higgs mass limits.}
\end{figure}

\begin{figure}[!bt]
  \centering \includegraphics[height=0.8\textheight]{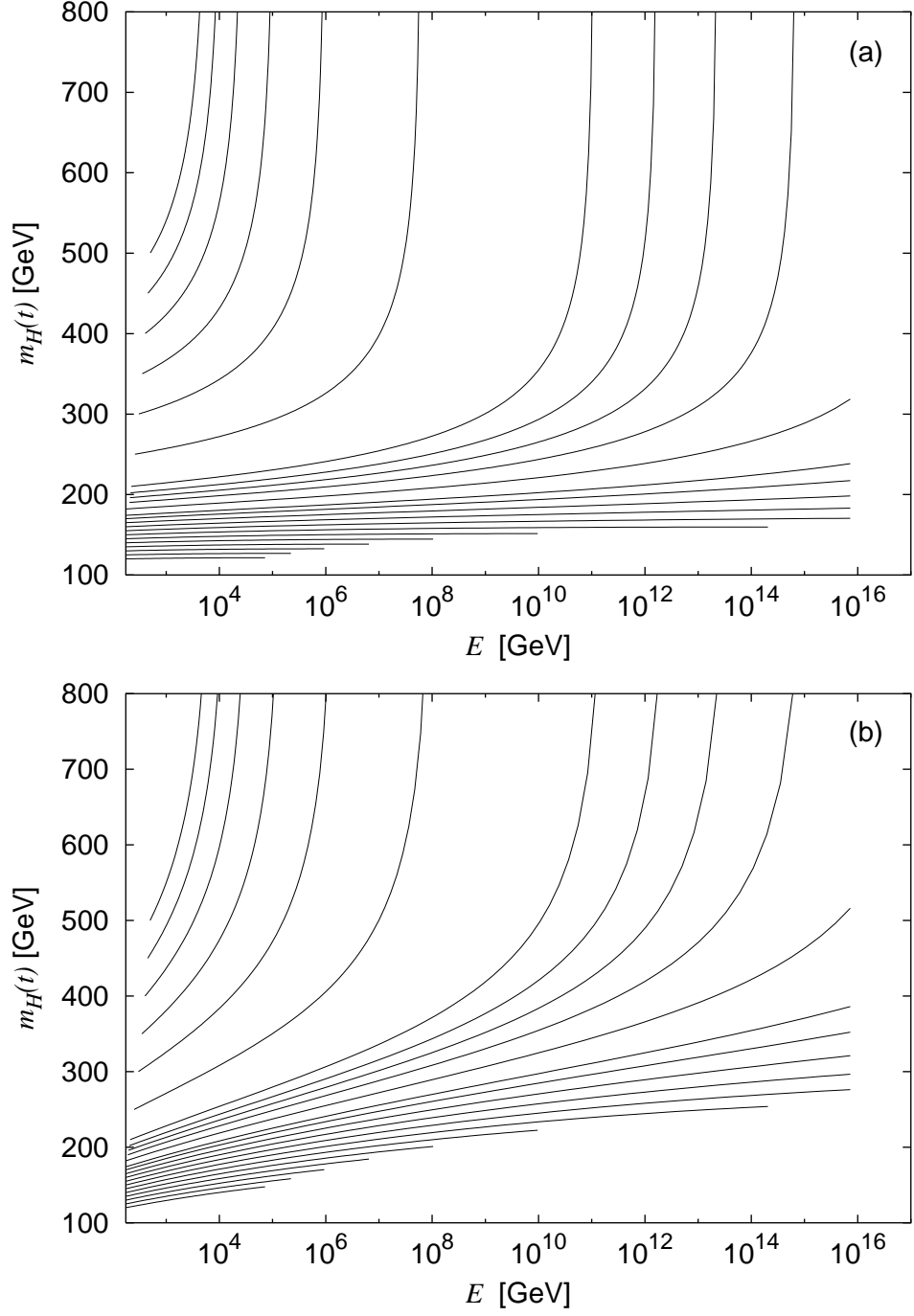}
\caption{\label{fig5}
  Running of the Higgs mass. Figure~(a) corresponds to the case of the
  effective potential and figure~(b) to the tree-level relation for
  the Higgs mass. In case~(a) for the Higgs masses $m_{H}<174$~GeV the
  function $m_{H}(t)$ has no singularity and is almost linear. For
  Higgs masses $m_{H}>190$~GeV $m_{H}(t)$ has a singularity (a pole)
  which defines the upper limit for the UV cut off. In case~(b) for
  the Higgs masses $m_{H}<164$~GeV the function $m_{H}(t)$ has no
  singularity and is almost linear. For Higgs masses $m_{H}>178$~GeV
  $m_{H}(t)$ has a singularity (a pole).  In both cases for the Higgs
  masses $\sim 500$~GeV $m_{H}(t)$ is a very steep function of $t$.}
\end{figure}

\begin{figure}[!bt]
  \centering \includegraphics[height=0.8\textheight]{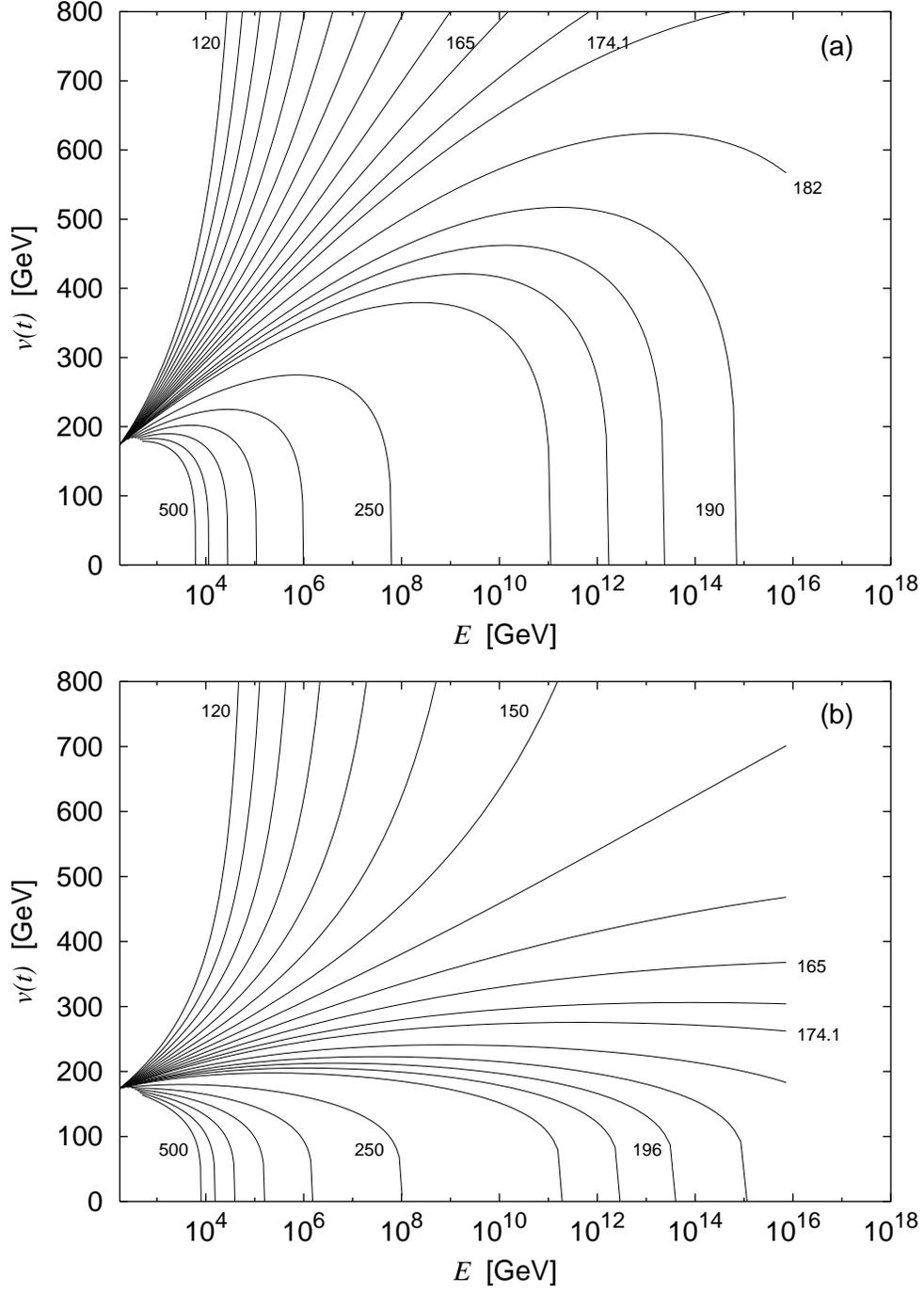}
\caption{\label{fig6}
  Running of the vacuum expectation value of the Higgs field $v(t)$.
  Figure~(a) corresponds to the case of the effective potential and
  figure~(b) to the tree-level relation for the Higgs mass.  The
  function $v(t)$ depends on the initial Higgs mass. The numbers at
  the ends of some plots correspond to the initial Higgs mass in GeV.
  In case~(a) for the Higgs masses $m_{H}<174$~GeV the $v(t)$ has a
  singularity at the UV cut off energy. For the Higgs masses
  $m_{H}>190$~GeV the function $v(t)$ for large values of $t$ is
  decreasing and at the UV cut off it vanishes. In case~(b) for the
  Higgs masses $m_{H}<164$~GeV the $v(t)$ has a singularity at the UV
  cut off energy. For the Higgs masses $m_{H}>178$~GeV the function
  $v(t)$ for large values of $t$ is decreasing and at the UV cut off
  it vanishes. The behavior of $v(t)$ in both cases is significantly
  different.}
\end{figure}

\begin{figure}[!bt]
  \centering \includegraphics[height=0.8\textheight]{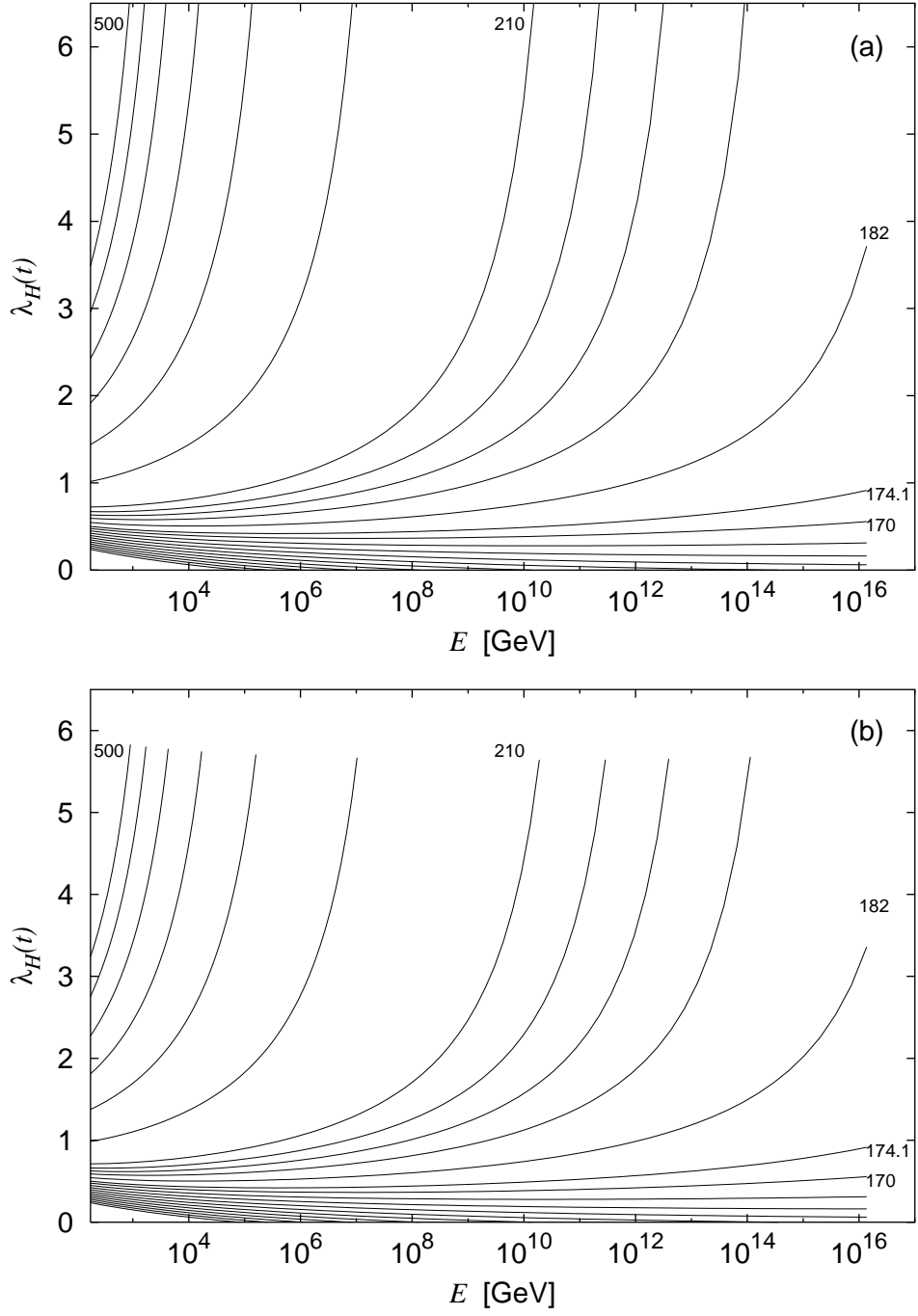}
\caption{\label{fig7}
  Running of $\lambda_{H}(t)$. Figure~(a) corresponds to the case of
  the effective potential and figure~(b) to the tree-level relation
  for the Higgs mass. The function $\lambda_{H}(t)$ depends on the
  initial Higgs mass. The numbers at the ends of some plots correspond
  to the initial Higgs mass in GeV. One can see that for low values of
  $\lambda_{H}(0)$ which correspond to the low Higgs masses the
  function $\lambda(t)$ has a zero and for higher values of
  $\lambda_{H}(0)$ corresponding to larger Higgs masses there appears
  a Landau pole. Both cases~(a) and~(b) give very similar predictions.}
\end{figure}

The numerical predictions for both cases are presented in
Figs.~\ref{fig4}, \ref{fig5}, \ref{fig6} and \ref{fig7}. These figures
contain the upper value for the UV cut off as a function of the Higgs
mass, running of the Higgs mass, VEV of the Higgs field and of
$\lambda_{H}(t)$. To obtain these figures we used the Higgs boson
matching scale equal to $m_{H}=\max\{m_{t},m_{H}(t_{H})\}$, where
$t_{H}=\ln(m_{H}/m_{t})$. The figures 4a, 5a, 6a, 7a correspond to the
case of the effective potential and the figures 4b, 5b, 6b, 7b
correspond to the case of the tree level case.

The left hand side of Fig.~\ref{fig4} consists of only one curve that
is obtained from the condition $\lambda_{H}(t)=0$. For the Higgs
masses that allow this condition there is no Landau pole up to the GU
energy $E_{GU}$ and the values of $\lambda_{H}$ are small and the
function $\lambda_{H}(t)$ is monotonically decreasing (see
Fig.~\ref{fig7}) so the two loop and higher corrections are small and
the perturbation series does not diverge.  The right hand side of the
figure consists of the three curves.  The upper curve corresponds to
the position of the Landau pole.  Since it is well known that for the
energies close the position of the Landau pole the perturbation series
breaks down and the two and higher order loop corrections become
important we have drawn two additional, more realistic curves that
correspond to the values of $\lambda_{H}$ when the two and three loops
corrections become important~\cite[~Eq.~(3.2),
$\lambda_{FP}\approx12.1$]{ref8}: the lower curve corresponds to
$\lambda_{H}=\lambda_{FP}/4$ and the middle curve to
$\lambda_{H}=\lambda_{FP}/2$. These curves differ significantly from
the one for the Landau pole curve only for the high Higgs masses.

From Fig.~\ref{fig4} we see that for the Higgs masses
$m_{H}\leq152$~GeV the UV cut off is growing as a function of the
Higgs mass. The energy of grand unification $E_{GU}$ is reached
at $m_{H}\approx150$~GeV and then there is a narrow window of the Higgs
masses
\begin{equation}
\label{22}
  \begin{split}
    &150 \leq m_{H} \leq 193\;\;\text{GeV}\quad
    \text{for the effective potential},\\
    &150 \leq m_{H} \leq 194\;\;\text{GeV}\quad
    \text{for the tree level mass relations},
  \end{split}
\end{equation}
for which the UV cut off exceeds the $E_{GU}$ scale. For the Higgs
masses $m_{H}\geq178$ there appears the Landau pole and the UV cut off
is decreasing as a function of the Higgs mass. Here, the most
realistic is the lowest curve in Fig.~\ref{fig4} which corresponds to
the point where the perturbation series ceases to be meaningful.

In Fig.~\ref{fig5} we show the evolution of the Higgs boson mass in
the range $120\leq m_{H}\leq500$~GeV. We see that according to the
initial Higgs mass there are two distinct patterns of evolution.  In
case~(a) of the effective potential for $m_{H}<174$~GeV there is a
slow linear evolution up to the UV cut off.  For $m_{H}>178$~GeV there
is a singularity at the UV cut off and the growth of $m_{H}(t)$ is
much faster especially for high Higgs boson masses. In case~(b) of the
tree level mass relations for $m_{H}<164$~GeV there is a slow linear
evolution up to the UV cut off.  For $m_{H}>178$~GeV there is a
singularity at the UV cut off and the growth of $m_{H}(t)$ is much
faster especially for high Higgs boson masses. The behavior of
$m_{H}(t)$ is significantly different for both cases.

In Fig.~\ref{fig6} we show the evolution of the VEV of the Higgs field
derived from Eq.~\eqref{eq21d3} for the Higgs masses in the range
$120\leq m_{H}\leq 500$~GeV. Similarly as in the case of the Higgs
mass there are two patterns of evolution. In case~(a) of the effective
potential for the Higgs masses $m_{H}<174$~GeV the VEV is a growing
function of energy and has a singularity at the UV cut off. For the
Higgs masses $m_{H}>182$~GeV the VEV initially increases with energy
then bends and hits zero at the UV cut off. In case~(b) of the tree
level mass relations for the Higgs masses $m_{H}<164$~GeV the VEV is a
growing function of energy and has a singularity at the UV cut off.
For the Higgs masses $m_{H}>178$~GeV the VEV decreases with energy and
hits zero at the UV cut off. The behavior of $v(t)$ is significantly
different for both cases.

In Fig.~\ref{fig7} we show the evolution of the coupling
$\lambda_{H}(t)$ given in Eq.~\eqref{17}. One can see that the
behavior of $\lambda_{H}(t)$ is in agreement with the earlier
discussion and for values of $\lambda_{H}(t_{0})<0.37$ the function
$\lambda_{H}(t)$ has a zero and for $\lambda_{H}(t_{0})>0.61$ it has a
pole. The dependence of $\lambda_{H}(t)$ on the Higgs mass for the
case~(a) of the effective potential and~(b) of the tree level mass
relations is very similar.

The behavior of the Higgs VEV has important implications for the
standard model. It means that the masses of the gauge bosons and
quarks would grow with energy for $m_{H}< 174$~GeV. For
$182<m_{H}<300$~GeV the masses of the gauge bosons and quarks would
first increase with energy, then reach a maximum and decrease. For the
Higgs masses $m_{H}>300$~GeV the VEV rapidly tends to zero.

From the evolution of the Higgs mass and VEV one can see the
appearance of two patterns of the high energy behavior of the standard
model. One, for $m_{H}< 174$~GeV and the other for $m_{H}> 182$~GeV.
The Higgs with $m_{H}=180$~GeV is the point of the transition between
the two patterns for the case of the effective potential and
$m_{H}=168$~GeV is the transition point for the tree level mass
relations.

It is interesting that both cases: the Higgs effective potential and
the tree level mass relations give very similar results for the Higgs
mass limits (see Fig.~\ref{fig4}) while the energy behavior of the VEV
and $m_{H}(t)$ is significantly different. This has the origin in the
fact that the position of the UV cut off obtained from Eqs.~\eqref{19}
and~\eqref{19a} is insensitive to the running of the Higgs mass
derived from Eqs.~\eqref{eq21d1} and~\eqref{kal3}.

The results of this paper support the results of
Refs.~\cite{ref8,ref8a} where the similar problem was considered. In
Ref.~\cite{ref8} the authors were using the simplified assumption that
the gauge couplings and the top quark Yukawa coupling are constant and
do not run according to the RGE. Our treatment is more precise and the
simplifying assumptions are not necessary. It is interesting that the
running of the gauge couplings and the top quark Yukawa coupling have
an important influence on the results especially for the low Higgs
masses, where the two loop corrections are negligible.

To conclude let us stress that the key point of the paper is the
treatment of the one loop RG equation for $\lambda_{H}$ and the
linearization of the problem by the substitution in Eq.~\eqref{eq9b}.
This linearization permitted very precise analysis of the positions of
the Landau pole and the point where $\lambda_{H}$ vanishes.  Moreover
it also gave an intimate relation between the positions of these two
points: Eq.~\eqref{19a} is the derivative of Eq.~\eqref{19}. Such a
relation between these two important quantities is a new result.
Additionally it should be also stressed that the \emph{analytical}
results for running of $\lambda_{H}$ up to one loop is a very good
starting for a precise analysis of the two loop effects.

\begin{acknowledgments}
  We are grateful to Drs.\ Marc Sher and Jos\'e Ram\'on Espinosa for
  drawing our attention to the method of the effective potential in
  the analysis of the Higgs mass. We thank Dr.\ Mikhail Kalmykov for
  the information about the Higgs mass relation and explaining its
  physical significance. We also thank Drs.\ Andrzej Buras and Gabriel
  L\'opez Castro for their suggestions. We acknowledge the financial
  support from CONACYT.  S.R.J.W.\ also thanks to ``Comisi\'{o}n de
  Operaci\'{o}n y Fomento de Actividades Acad\'{e}micas'' (COFAA) from
  Instituto Polit\'{e}cnico Nacional and thanks are also to Dr.~Paul
  Singer from the Technion in Haifa for his hospitality where part of
  the work was done.
\end{acknowledgments}


\end{document}